\documentclass[preprint]{aastex}
\usepackage{natbib}
\usepackage{psfig}

\shortauthors{McLaughlin and Wijers}
\shorttitle{Nickel Decay in GRBs}

\newcommand{\grb}{$\gamma$-ray burst}
\newcommand{\grbs}{$\gamma$-ray bursts $\,$}

\begin{document}


\title{Delayed Nickel Decay in \grbs}


\author{G. C. McLaughlin}
\affil{Department of Physics, North Carolina State University, Raleigh, NC 
27695-8202}
\email{Gail\_McLaughlin@ncsu.edu}

\and

\author{R. A. M. J. Wijers}
\affil{Department of Physics and Astronomy, State University of New York,
    Stony Brook, NY 11794-3800}
\email{Ralph.Wijers@sunysb.edu}

\begin{abstract}

Recently observed emission lines in the X-ray afterglow of gamma ray bursts
suggest that iron group elements are either produced in the \grb \, or
are present nearby. If this material is
the product of a thermonuclear burn, then such material would be expected 
to be rich in Nickel-56.  If the nickel remains partially ionized, this
prevents the electron
capture reaction normally associated with the decay of Nickel-56, 
dramatically
increasing the decay timescale.  
Here we examine the consequences of rapid ejection of
a fraction of a solar mass of iron group material from the center of
a collapsar/hypernova.  The
exact rate of decay then depends on the details of the ionization and
therefore the ejection process.  Future observations of
iron, nickel and cobalt lines can be used to diagnose the origin
of these elements and to better understand the astrophysical site of
\grbs. In this model the X-ray lines of these iron-group
elements could be detected in suspected hypernovae that did not produce
an observable \grb\ due to beaming.
\end{abstract}



\clearpage
\section{Introduction}

For the last thirty years
\grbs have been observed,  but it has only been in the
last five years that X-ray, optical, and radio counterparts ---the 
\lq afterglows'---
have been seen \citep{Costa97B,Paradijs97,Frail97A}.  
Pinpointing the location of the afterglows has allowed
some of the bursts to be associated with host galaxies.  Absorption lines
in the optical afterglow, together with redshifts obtained from the
host galaxies have allowed redshifts to be obtained for the \grbs 
themselves, e.g. \citep{Metzger97B}.  
Redshifts have been obtained between $z = 0.35$
and $z = 4.5$, confirming the idea that bursts are cosmological in origin.
This idea is also supported by their isotropic distribution
on the sky, as recorded by over 2700 BATSE detections.  

The redshift determinations have led to estimates of isotropic equivalent 
fluxes in $\gamma$ rays, ranging from $10^{50} {\rm ergs}$ to 
$10^{54} \, {\rm ergs}$.  Since the upper bound is approaching the
rest mass energy of the sun, the astrophysical origin
of these sites is fairly restrictive.  
Less energy is contained in the bursts if the
$\gamma$ rays are strongly beamed, but the beaming can not be stronger
than a few degrees, based on the lack of detection of 
\lq homeless afterglows'
\citep{Meszaros99A,Dalal02,Vreeswijk02,VandenBerk02}.  
Recent analyses of the break in
the afterglow signal support the idea that \grbs are beamed to a few degrees
\citep{Frail01,Kumar01}.

The theoretical problem of what causes the \grbs \, can be 
divided has several parts: the fireball model, the energy injection
mechanism and the astrophysical site.  
The mechanism for actually producing the $\gamma$ rays and the afterglows
is described by the relativistic fireball model  
\citep{Rees92,Meszaros97,Wijers97B}.  After energy is injected into 
material composed mostly of 
electrons and positrons, it is ejected relativistically.  
Internal shocks in the plasma and
external shocks caused by contact with the interstellar medium drive 
synchotron radiation from the electrons in a magnetic field.  The
spectrum of the afterglow, at least in some cases, is fairly well fit
to a synchotron spectrum.  For recent reviews see 
\citealt{Paradijs00,Piran99}.  

This model is fairly independent of the initial energy injection
mechanism, as well as the astrophysical site for the bursts.  For the
injection mechanism, neutrino and antineutrino annihilation into 
electron-positron pairs (e.g., \citealt{Ruffert99}), 
and delivery through a pointing flux such as the Blanford-Znajek mechanism,
have been discussed (e.g., \citealt{Brown00B}).  

Two
contenders for the site of \grbs are neutron star - neutron
star mergers and \lq failed supernovae' also known as collapsar or hypernova 
models.  The neutron star models \citep{Eichler89,Mochkovitch93}
have been shown to be dependent on the
way in which general relativity is handled in the numerics.  Janka (1999),
using a post newtonian approximation, finds short bursts with at most
$10^{50}$ ergs ejected.  Salmonson et~al.\ (2001) using a GR approximation 
that is exact in the case of spherical symmetry, find longer timescales
and larger energies.  The collapsar models involve the collapse of a massive
star which fails to produce a viable shock for an ordinary supernova
explosion.   Models involve a combination of a 
rotating black hole and magnetic field
driving jets along the rotation axis of the black hole  
\citep{Woosley93B,Paczynski98,MacFadyen99}.
Of these two models, the collapsar model has recently gained favor with
identification, in many cases, of host galaxies associated with the
\grb\ \citep{Bulik99,Bloom99A}.  Neutron star binaries would be expected 
to wander out of 
galaxies at a rate inconsistent with the number of observed associations
with host galaxies. Further, in a number of \grb\ light curves, features
strongly resembling a supernova have been found.

Furthermore, recent observations of K $\alpha$ lines from iron have been 
identified in the X-ray afterglows of four \grbs \citep{Piro99A, Antonelli00,
Piro00B}.  This is more likely to occur in a supernova like model than 
in the neutron star merger
models.  However,  there are still many unsolved puzzles associated 
with these lines.  The lines are too broad (0.5 KeV) to be accounted
for by thermal broadening, although this may be accounted for
by electron scattering. Line scattering from reprocesser type
models is discussed in, e.g., \citep{Weth00, McLaughlin01, Kallman01}.
If the lines come from
material produced by nucleosynthesis in a \grb, then one would expect
Nickel-56 as the product of a thermonuclear burn.  In most cases
the lines
have been identified as iron, in some cases with independent
redshift measurements.  However, the most recent detection is of nickel,
not iron \citep{Reeves02}.

Two general classes of models have been proposed for these lines.  In 
one case the iron is already present in the system before the $\gamma$
rays appear and the lines are produced by the interaction of the afterglow
with the iron, e.g., the Supranova model \citep{Vietri01}.  In the other
case the lines are produced by another mechanism and are not
directly associated with the afterglow.  For example, one might
imagine that the iron is present in what is left of the star after
the burst.  An accretion disk which has formed around the black hole
produces an ionizing flux, that reaches this iron, which sits on the 
surface of a \lq funnel \rq\ which has been carved out of the star by the
original ejection, (cf. \citealt{Rees00}). 

Nickel, on the other hand,  is likely to be produced by
a thermonuclear burn in the silicon burning layer
of the star.  This could be identified by a time dependence in
the energy of the observed line, due to the decay from
nickel to cobalt to iron \citep{McLaughlin01}.  
This could be principle be detected up
to a few days after the event, since the half life for nickel is
about six days, and the difference in line energy from nickel to
cobalt is about half a keV.  An alternative mechanism
suggested for the production of nickel is that first material in 
the very hot accretion disk is completely dissociated into protons and
neutrons. A wind from this disk will cool as it
moves out, and may recombine to an equilibrium nucleus, likely again
nickel, \citep{MacFadyen02}.

In this paper we suggest a variation on the above scenarios.  
Nickel, produced by a thermonuclear burn in the silicon shell, 
may be ejected out of the
accretion disk surrounding the black hole in the collapsar model.
This nickel will be partially ionized from below, by the flux
leaving the accretion funnel. The ionized nickel
will decay at a different rate than nonionized nickel, since
decay of nickel proceeds by electron capture.  This will produce a
distinctive signature pattern in the time dependence of
recombination lines.  In section \ref{sec:model} we make a plausibility
argument to show that ejected nickel can remain ionized.  We discuss the
parameters that determine the optical depth of the
ejected nickel. We describe the
progress of an ionization front through the ejected nickel and
its dependence on various parameters.  In section 
\ref{sec:nuclear} we discuss the relevant nuclear physics of 
electron capture and beta decay of nickel-56 and cobalt-56.
In section \ref{sec:results} we give the fractions of these elements
as functions of time in this model.  In section \ref{sec:conclusions}
we give conclusions.

\section{The Model}
\label{sec:model}

In this section we present a model of material ejected from a 
the center of a \grb.  We assume that this occurs by way
of interaction between the rotating accretion disk surrounding
a black hole and the magnetic field.  
If the \grb \, is a collapsar/hypernova, then
the material is likely to be rich in nickel-56.  This will happen 
if the silicon burning shell with an electron fraction of roughly 
0.5 is heated to $T > 4\times10^9$\,K.  This
type of burning occurs in ordinary supernova explosions.
Some of this nucleosynthesis may be
deposited in the accretion disk. 

First we discuss the optical depth of the nickel and then we describe 
the motion of an ionization front which travels through the ejected material.

\subsection{Optical Depth}
In this subsection we describe the optical depth of 
 material ejected with fairly high velocity 
($\beta \approx v/c \approx 0.2$) 
from the center of a \grb.  There is an ionizing flux
coming from the center of the object, presumably from an accretion disk
surrounding a black hole.  As the material moves away from the center
it also expands, so we approximate its size as roughly 
$V \sim \Omega r^2 \Delta r$, where 
$r = \beta c t $,  $\Omega = 2 \pi (1 - \cos \theta)$ and 
$\Delta r \approx r$.  Here $r$ is the linear dimension, $t$ is time and 
$c$ is the speed of light. For definiteness, we assume in our calculations
that the expansion is homologous, i.e., $v\propto r$, and the leading edge
moves with $v=\beta c$. This way, the density of the ejecta is independent
of position, and scales with time as $t^{-3}$. Since the ejecta are highly
supersonic, and the radiation force is small except near the inner edge,
it is fair to approximate the velocity of each part of the ejecta as
constant.

We wish to determine how much of the nickel is fully ionized, as both a 
function of position and time.  The first step is to estimate the  
recombination rate and the ionization rate for the expanding
material. The ionization rate is given by:
\begin{equation}
R_{ion} = \int_{\nu_Z}^{\infty} \sigma(\nu,Z) {F(\nu)  \over h \nu} {\rm d}\nu
\end{equation}
We use $\sigma(Z) = 2.8 \times 10^{29} Z^4 / \nu^3$,
for the ionization cross section.  The lower limit of integration is the
binding energy of the final K shell electron, $ h \nu = 13.6 Z^2 \, {\rm eV}$. 
$F(\nu)$ is the photon energy flux.

This flux may have different forms depending on the details of the model.
The observed ionizing fluxes are much larger than can be supplied by 
Eddington-limited emission from a disk around a black hole of a few solar
masses. However, a mechanical and/or poynting wind generated by the
Blandford-Znajek mechanism or neutrino annihilation can have a much larger
energy flux, and be converted into heat and radiation when it impacts the
funnel wall far from the black hole (or from a highly magnetized
neutron star), as envisaged by Rees \& M\'esz\'aros (2000). In that case,
the expected spectrum might be more like synchrotron radiation or 
bremsstrahlung with a very high temperature; in both cases, we may approximate 
the keV X-ray spectrum with a power law:
$F(\nu) \propto \nu^{-\xi}$.

This is related to the number flux of the material
above the ionization threshold as
\begin{equation} 
\dot{N}_I  =  4 \pi r^2 \int^{\infty}_{h \nu_Z} {F(\nu) \over h \nu}.  
\end{equation}
We parameterize the time dependence of the ionizing source 
(in $10^{53}$ photons per second) as
\begin{equation}
\dot{N}_I = N_{I,53} t_d^{-\alpha} 10^{53} {\rm s}^{-1}
\end{equation}    
where time, $t_d$ is measured in days.
After some algebra, we rewrite the ionization rate as
\begin{equation}
R_{ion} = 6.5 \times 10^ 6 t_d^{-(2+\alpha)} N_{I,53} 
\left( {0.2 \over \beta} \right)^2 
\left( {28 \over Z} \right)^2 \zeta,
\end{equation}
where
\begin{equation}
\zeta = {\int^{\infty}_{\nu_Z} {F_\nu \over h\nu} {1 \over \nu^3} d \nu 
\over \int^{\infty}_{\nu_Z} {F_\nu \over h\nu} d \nu},
\end{equation}
i.e., it is a form factor expressing the ionizing photon number flux with 
the cross section above the edge threshold. For a power-law spectrum 
$F(\nu)\propto\nu^{-\xi}$, we have 
$\zeta ={4 \xi \over \xi + 3}$.

Also necessary is the recombination rate, which we
estimate by starting with the
recombination luminosity \citep{Lang80},
\begin{equation}
L_{rec} = 10^{-21} n_e n_i T^{-1/2} V Z^4 \; {\rm erg} \; {\rm s}^{-1}
\end{equation}
Here temperature is in Kelvin, electron, $n_e$  and ion, $n_i$ 
densities in ${\rm cm}^{-3}$, and V is volume.  The recombination
rate per ion can be estimated by dividing by the volume and 
ion density and also by the energy of the photon,
$h \nu \approx Z^2 13.6 {\rm eV}$.  Some algebra yields a
recombination rate of
\begin{equation}
R_{rec} = {C_{rec} \over t_d^3} {\rm day}^{-1}
\end{equation}
with time t in days where
\begin{equation}
C_{rec} = 1.1 \times 10^6 \left( {1 {\rm keV} \over T } \right)^
{1/2}
\left( { M \over 0.1 M_\sun} \right) 
\left( {0.2 \over \beta} \right)^3
\left( {1 - \cos 20^\circ  \over 1 - \cos \theta} \right) 
\left( { Z \over 28 } \right)^3
\left( { A \over 56} \right).
\end{equation}
Here M is the amount of mass in the ejected material and $\theta$
is the opening angle of the cone. $T$ is the temperature of the
material.

With the recombination and ionization rates we can calculate the
fraction of nonionized nickel in the optically thin
region of the material as
\begin{equation}
\label{eq:fnon}
f_{non} = {1 \over 1 + C_{i} t_d^{1 - \alpha}}
\end{equation}
where the coefficient is
\begin{equation}
C_{i} \approx 6 
\dot{N}_{I,53}
\left( {1 - \cos \theta  \over 1 - \cos 20^\circ} \right) 
\left( {\beta \over 0.2} \right)
\left( {.1 M_\sun \over M} \right)
\left( T \over {\rm keV} \right)^{1/2} 
\left( {28 \over Z} \right)^5
\left( {56 \over A} \right)
\zeta 
\end{equation}
If the luminosity of the ionization source is constant, then
the fraction of nonionized nickel decreases with time. This may be
counterintuitive, since the material gets further from the ionizing source.
But the ionization rate per atom decreases with flux, i.e., as $1/r^2\propto
1/t^2$ (since $r\propto t$),
whereas the recombination rate is proportional to density, i.e., scales
as $1/r^3\propto 1/t^3$, hence ionization wins.
However, in case of an emptying accretion disk, or of a spinning-down 
black hole, the luminosity will scale roughly as $t^{-1}$, which implies
a roughly constant ionized fraction.

The optical depth of the ionizing photons, $\tau = f_{non} n_i \sigma_Z r $ is
an decreasing function of time.  Therefore an ionization front passes through
the material.  The details of this ionization front determine the
ratio of ionized to nonionized nickel, in the material and therefore the 
rate of nickel decay.    

\subsection{Ionization Front}

We assume that Nickel 56 is ejected from the accretion disk with fairly
high velocity.  There is a source of ionizing photons, which emits
in photons per unit time,
\begin{equation}
\dot{N} = N_{I,53} 10^{53} /t_d ^\alpha \; {\rm s}^{-1} = C_s N_i  / t_d^\alpha
\; {\rm day}^{-1}, 
\end{equation}
where $N_i$ is the total number of ions in the material 
and 
\begin{equation}
C_s = 4 \times 10^3 
\left( {M \over .1 M_\sun} \right)^{-1} \left( {A \over 56} \right) N_{I,53} .
\end{equation}    

The equation which describes the passage of the ionization front is:
\begin{equation}
\label{eq:ionfront}
N_{I,53} 
{1 \over t^\alpha} \Delta t = 
n_i f_{non} R_{ion} \Omega r^3 r_{f_i}^3  \Delta t 
+ n_i \Omega r^3 r_{f_i}^2 \Delta r_{f_i}
\end{equation}
The photons are either absorbed on their way to the front (second term)
or at the front (first term).
Here $r$ is the linear dimension of the ejected material, and 
$r_f = r / r_{max}$ is the fractional distance within that material,
while $r_{f_i}$ is the position of the ionization front.
The solid angle subtended by the ejected material is 
$\Omega = 2 \pi (1 - \cos \theta)$.  We rewrite 
Eq. \ref{eq:ionfront} as
\begin{equation}
{3 C_s \over t_d^\alpha} = {3 C_{rec} \over t_d^3} r_{f_i}^3 + 
{r_{f_i}^3 \over 3}{d r_{f_i} \over dt}
\end{equation}
The solution to this equation is
\begin{equation}
\label{eq:rfi}
r_{f_i} \approx \left( C_s \over C_{rec} \right)^{1 \over 3} 
t_d^{ 3 - \alpha \over 3},
\end{equation}
as long as $(3 C_{rec} / 2)^{1/2} >> t_d$ during the time the ionization
front is passing through the material.  This is true for all the 
situations considered here.      

Eq. \ref{eq:rfi} describes the motion of the ionization front.  The
time it takes a front to pass through the entire mass of nickel depends
strongly on $\alpha$.  For example, for the parameters considered here, it
takes 6 days for the front to pass through the material if 
$\alpha = 0$ and 17 days for the front to pass through if $\alpha = 1$.

We have now described a moving, expanding mass of nickel, which
is partially ionized behind the front and not ionized ahead of the front.
We wish to determine the decay properties of the nickel, for that we
need the nuclear physics described in the next section.

\section{Electron Capture and Beta Decay}
\label{sec:nuclear}
Here we describe the nuclear physics of the A=56 decay chain.
The half life of $^{56}{\rm Ni}$ is 6.075 days, and it 
decays almost exclusively by electron capture,  
\begin{equation}
^{56}{\rm Ni} + e^- \rightarrow ^{56}{\rm Co} + \nu_e.
\end{equation}
The Q-value of this reaction is
2.136 MeV. The decay proceeds primarily through the 3rd and 4th excited
states of Cobalt-56 which are at 1.45 Mev and 1.720 Mev with respect 
to the ground state.  These states have spin and parity of $0^+$ and 
$1^+$,
while the ground state of nickel is $0^+$, so the decay proceeds through
Gamow-Teller and Fermi transitions.  

If the nickel is completely ionized and in a dilute environment
(so free electron capture is negligible), then it can
only decay by emission of a position,
\begin{equation}
^{56}{\rm Ni}  \rightarrow ^{56}{\rm Co} +e^+ +  \nu_e.
\end{equation}  
The Q-value for this beta plus decay is 1.114 MeV, which makes the 3rd and
4th excited states in cobalt energetically inaccessible.  
There are three remaining possibilities.  The decay can proceed through
the ground state, $4^+$, 	
the first excited state at 0.158 Mev, $3^+$ or the
second excited state at 0.970 MeV at $2^+$. These are all forbidden 
and to date no $\beta^+$ decay from Nickel-56 has ever been seen.  The 
current experimental limits place the half life at greater than 
$2.9 \times 10^4 {\rm years}$ \citep{Sur90}.  
Calculations within the large scale shell model 
have placed the half life at about $4 \times
10^4 \, {\rm years}$ 
\citep{Fisker99}.  For our 
purposes, we assume that completely ionized nickel is quasi-stable.  
We note that because of the long half life of ionized nickel-56, it
has been suggested that it may be seen in cosmic ray detectors.

Cobalt-56 also decays primarily (81\%) by electron capture into Iron-56.
\begin{equation}
^{56}{\rm Co} + e^- \rightarrow ^{56}{\rm Fe} + \nu_e.
\end{equation}  
Its half life is
77.2 days \citep{Junde99}.  The Q-value, 4.566 MeV, is much higher 
than in the case of nickel, and therefore it has more 
energetically accessible states in iron into which it may decay.  Nineteen
percent of the time ($I_{\beta^+} = 0.19$), 
Cobalt-56 decays by emission of a positron,  
\begin{equation}
^{56}{\rm Co} \rightarrow ^{56}{\rm Fe} + e^+ + \nu_e.
\end{equation}
the majority of which is  a $4^+$ to $4^+$ 
transition to the second exited state of iron at 2.085 MeV.  This decay will
still proceed, even if the iron is fully ionized.  
Therefore, for ionized cobalt, the
half life will be roughly a factor of 5 higher than for cobalt with two or 
more electrons.  Recent data on Cobalt-56 decay can be found in 
\citep{Junde99,Meyer90}.  
       
\section{Results: Nickel decay in \grb \, ejecta}
\label{sec:results}
We now combine the physics of the previous two sections to describe
 the motion of
the ionization front with the differing decay times of ionized and
nonionized nickel and iron.
We show that this produces unusual decay patterns that may be
observable with an instrument such as the Chandra X-ray observatory.

The rate of change of the relative abundances of 
Nickel-56, Cobalt-56 and Iron-56
\begin{equation}
{dN_{Ni} \over dt}  = -f_{non} {N_{Ni} \over \tau_{Ni}}
\end{equation}
\begin{equation}
{dN_{Co} \over dt} = f_{non} {N_{Ni} \over \tau_{Ni}}
-f_{non} {N_{Co} \over \tau_{Co}}
- (1-f_{non}) I_\beta^+ {N_{Co} \over \tau_{Co}}
\end{equation}
\begin{equation}
{dN_{Fe} \over dt} = f_{non} {N_{Co} \over \tau_{Co}}
+ (1-f_{non}) I_\beta^+ {N_{Co} \over \tau_{Co}}
\end{equation}
Here $\tau_{Ni} = \tau_{1/2} \ln 2$ and $\tau_{1/2}$ is the half life of
nickel and we use the similar relation for cobalt.  Before the ionization
front arrives, when $t < C_{f_\alpha} r_f^{3/(3 -\alpha)},
C_{f_\alpha} = (C_{rec} / C_{s})^{1 /(3 - \alpha)} $ the 
nonionized fraction is essentially unity.  After the ionization front
passes, the nonionized fraction is given by Eq. \ref{eq:fnon}.
We approximate the
nonionized fractions of nickel, cobalt and iron as the 
 same.  In reality, there will be slight quantitative differences, 
since the ionization potential changes by roughly a keV from nickel to iron. 
How this difference translates into $f_{non}$ depends
on the shape of the spectrum of the
ionizing source.

We can solve the above equations approximately analytically, to get the
amount of nickel which remains as a function of time and also
position in the ionization front.  The general expression for the fraction
of nickel remaining after the front has passed is,
\begin{equation}
\label{eq:niaf}
N(r_f,t) = \exp \left(-{C_{f_\alpha}r_f^{3 \over 3 - \alpha} \over \tau_{Ni}} \right)
\exp\left(- {C_{non}  \over \tau_{Ni}} 
\int^t_{C_{f_\alpha}r_f^{3 \over 3 -\alpha}} {dt \over 
C_{non} + t_d^{1 - \alpha}}\right), \hspace*{1cm} t_d > C_{f_\alpha} r_f^{3 \over 3 - \alpha}
\end{equation}
Before the front has passed,
\begin{equation}
N(r_f,t) = \exp(-t_d / \tau_{Ni}), \hspace*{1cm}  
t_d < C_{f_\alpha} r_f^{3 \over 3 - \alpha}.
\end{equation}

For the special cases of $\alpha = 0$ and $\alpha = 1$, Eq. \ref{eq:niaf}
becomes,
\begin{equation}
N(r_f,t_d) \approx \exp(-C_{f_0} r_f /\tau_{Ni}) \left( {C_{non} + 
   t_d \over C_{non} + C_{f_0} r_f} \right)^{-C_{non} / \tau_{Ni}}, 
     \hspace*{0.5cm} t_d > C_{f_0} r_f \hspace*{0.5cm} \alpha = 0
\end{equation}
\begin{equation}
N(r_f,t_d) \approx \exp(-C_{f_1} r_f^{3/2} / \tau_{Ni}) 
\exp \left[ { -f_{non} \over \tau_{Ni}} \left( t - C_{f_1} r_f^{3/2} \right) 
  \right], \hspace*{0.5cm} t_d > C_{f_1} r_f^{3/2} \hspace*{0.5cm} \alpha = 1.
\end{equation}

\begin{figure}
\vspace*{-4cm}
\begin{center}
\begin{minipage}{8.5cm}
\psfig{file=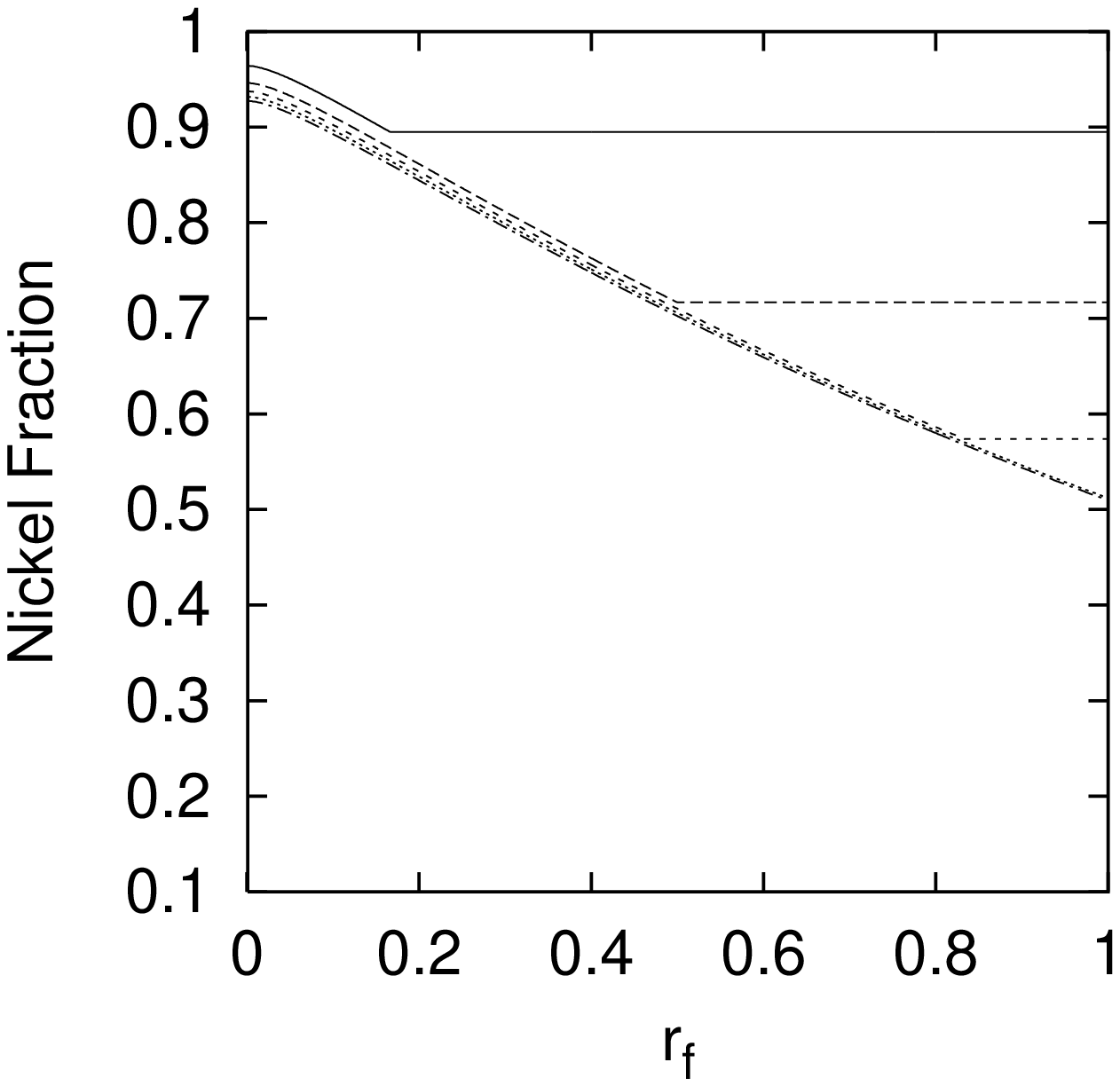,width=8.5cm}
\end{minipage}\begin{minipage}{8.5cm}
\psfig{file=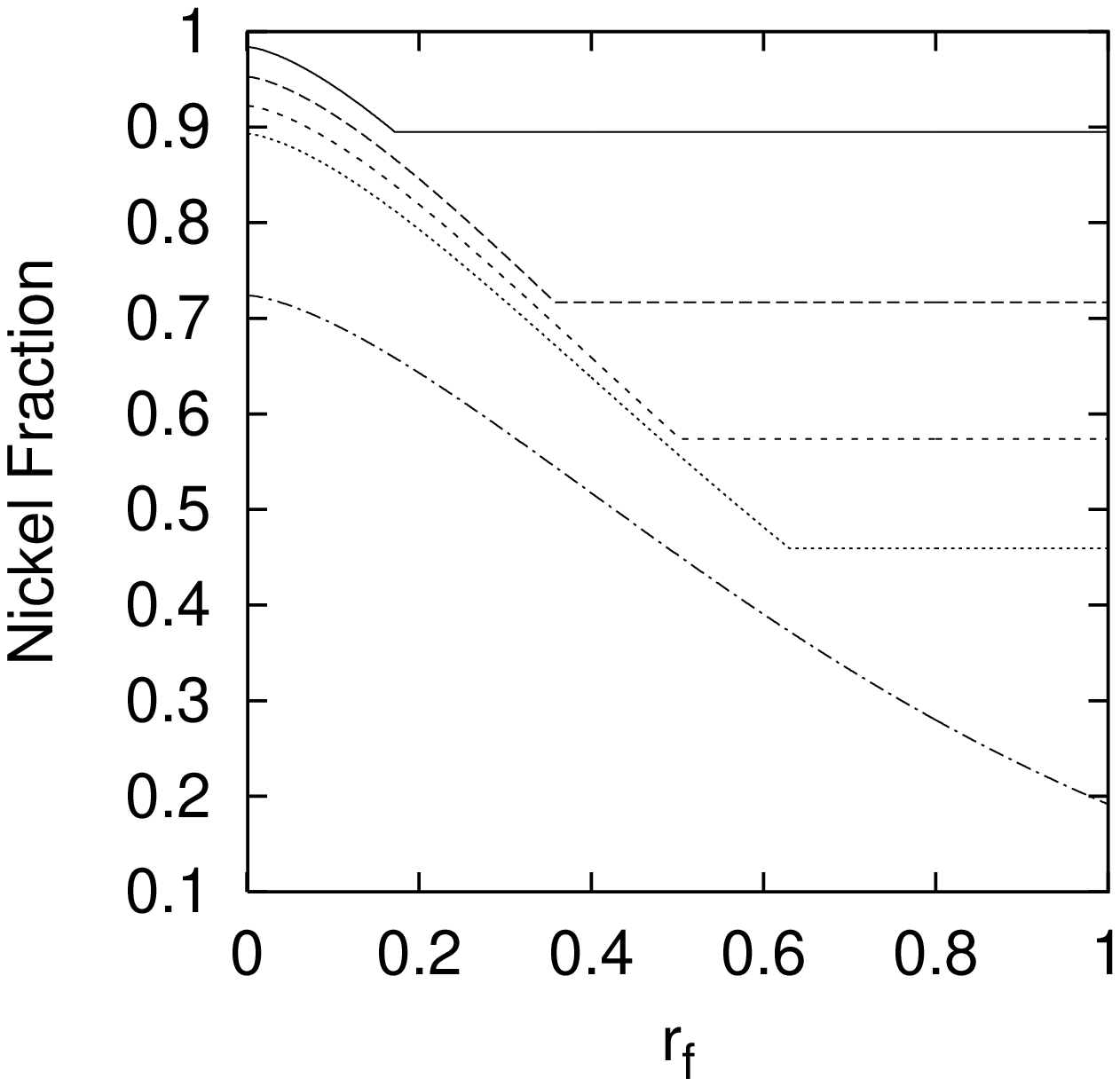,width=8.5cm}
\end{minipage}
\end{center}
\caption{
Fractional abundance of nickel as a function of position of the expanding 
material.  The horizontal axis represents relative position, zero nearest 
the ionizing source and one at the furthest edge.    
The different curves represent different times since the initial
ejection.  The top curve is after one day and the rest are, in decending order
3 days, 5 days, 7 days and 20 days.  Ahead of the ionization
front, the material decays normally, as seen by the horizontal lines.  
In the left panel the number of photons released from the ionizing 
source is constant as a function of time. In the right panel the ionizing 
source decays inversely with time, $\dot{N} \propto t_d^{-1}$.
\label{fig:alpha0}}
\end{figure}

\begin{figure}
\vspace*{-4cm}
\begin{center}
\psfig{file=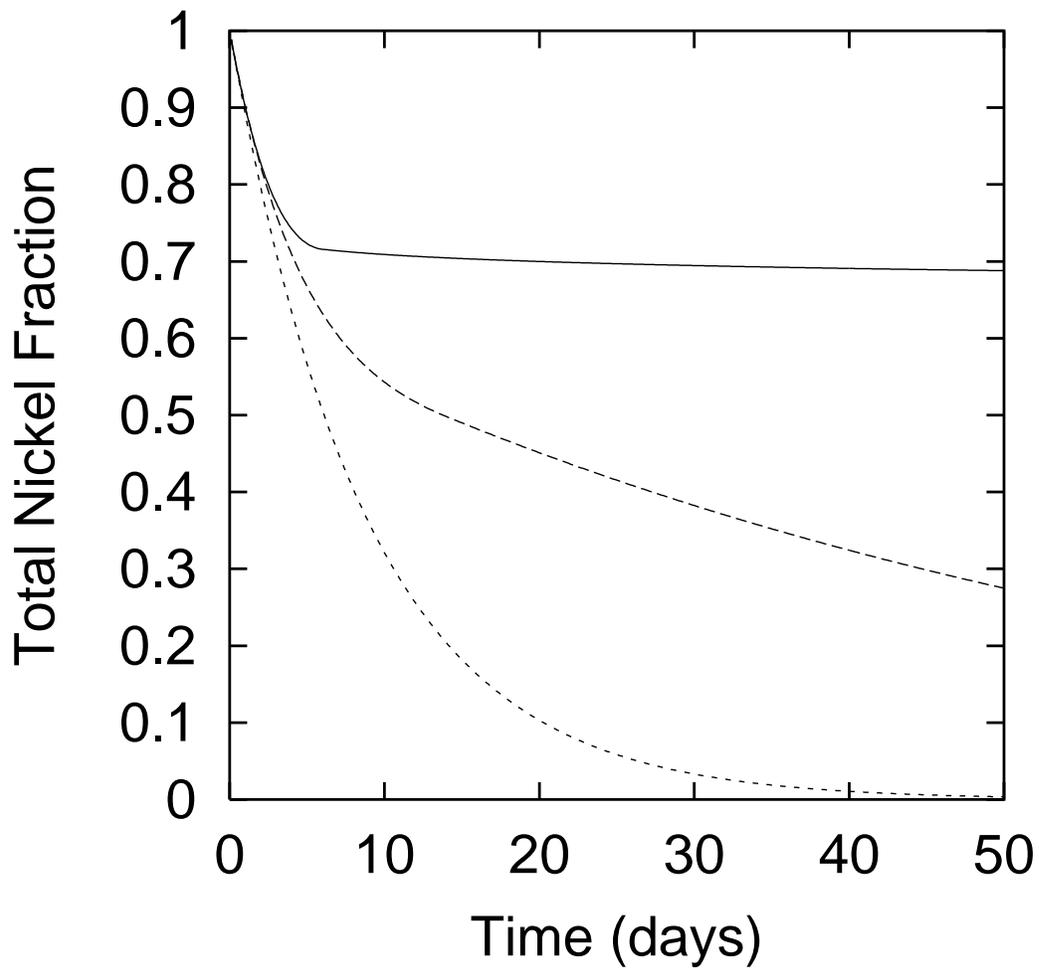,width=0.9\textwidth}
\end{center}
\caption{ This figure plots the total fractional amount of nickel in
the ejected material as a function of time.  The top curve is for a
constant ionizing source  while the second is for a 
source which decays inversely with time, $\dot{N} \propto t_d^{-1}$.  
The lowest curve
shows the fraction of nickel in nonionized material.  
\label{fig:nickeltot}}
\end{figure}

We show the results of our calculations in 
Figs.~\ref{fig:alpha0} and 
\ref{fig:nickeltot}.  In Figs \ref{fig:alpha0}a,b 
we show the fraction of nickel as a function of position, $r_f$, in the
ejected material, for various times.  One can clearly see the progress of 
the ionization front.  Ahead of the front there is a constant rate of
decay as indicated by the horizontal line.  Behind the front
the material is partially ionized and this slows down the rate of
decay.  In the case where the flux from the accretion disk remains
constant, the decay is much slower than in the case where the flux 
decreases as $t_d^{-1}$.

In Fig. \ref{fig:nickeltot} we show the total fraction of nickel
integrated over position.  The lowest curve shows the amount of nickel
as a function of time in the case of completely nonionized nickel.
In the upper two curves we show the change with rate of decay of the
ionizing source.

\begin{figure}
\vspace*{-4cm}
\begin{center}
\psfig{file=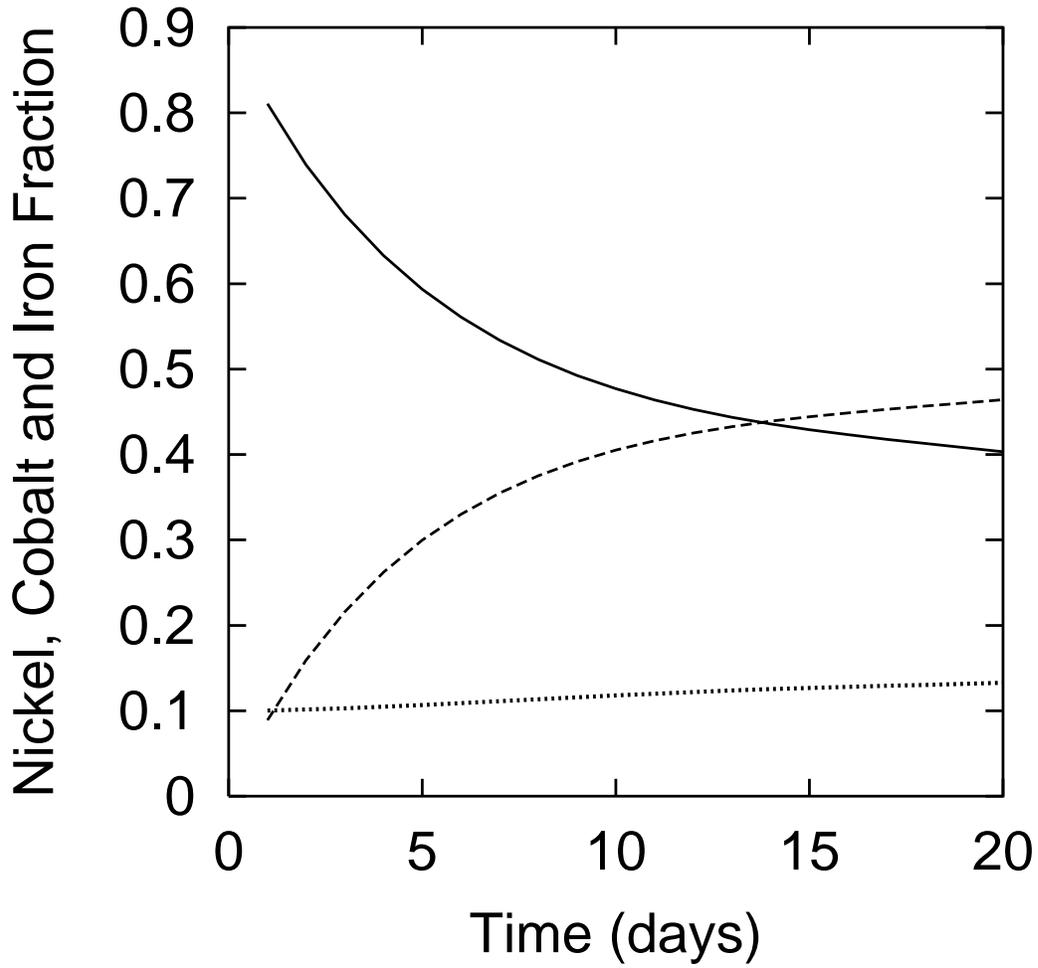,width=0.9\textwidth}
\end{center}
\caption{ This figure plots the total fractional amount of nickel (solid), 
cobalt (dashed) and iron (dotted)
as a function of time, for the parameters given in the
text and $\dot{N} \propto t_d^{-1}$.
\label{fig:irongroup}}
\end{figure}

In Fig. \ref{fig:irongroup} we show the amounts of nickel, cobalt and
iron as functions of time.  We begin with relative abundances taken
from \citep{Woosley95B}, so that the iron group has roughly
90\% Nickel-56 and 10\% iron.  We then follow the decay at every point in
the ejected material as the ionization front passes and we integrate over
position to look at the total amount of nickel, cobalt and iron 
as functions of time.  In these examples, the ionization front passes
through the ejecta in 1--2 weeks. Strictly speaking that poses a problem
for seeing the iron group emission lines reported in some afterglows
within a day, since it is presumed that we see \grbs\ along the direction
of the funnel and the ejecta. In that case, we cannot see a significant line 
flux along the jet direction until the ionization front has passed through
the ejecta. Real life, however, is likely to be more complex, with
homogeneities in the ejecta and a dependence of properties on the distance
to the jet axis, allowing the line photons to escape in some directions
in some bursts. Alternatively, if the ejecta start out hot enough, as may
be the case in a neutrino-driven disk wind, they may fully never recombine
and always be more optically thin.
Indeed, the lines are not seen in all afterglows and are
seen only part of the time in those afterglows where they have been seen.

\section{Conclusions}
\label{sec:conclusions}

The astrophysical origin of \grbs  remains a mystery even three decades
after their discovery.  However, technological advances in observing techniques
have resulted in a great increase in data which is helping narrow down 
the possibilities.  The recent observations of iron recombination lines in
the X-ray afterglows of \grbs \, are an important clue to this problem.

Here we have discussed the importance of potentially observing the
decay timscale of iron group lines.  The rate of decay will reflect
the amount of time newly made Nickel-56 has remained ionized.  This
information must then be interpreted in the context of the astrophysical
site for \grbs.

We have suggested a model which has Nickel-56 ejected from an accretion
disk surrounding a black hole. Such a geometry is
likely to exist at the center of a \grb.
We have shown that the ejected material remains ionized in part due to the
ionizing flux coming from the accretion disk.  The degree of 
ionization depends on the parameters in the problem
such as the ejection velocity and importantly, the rate of decay of the
flux coming from the accretion disk.  We suggest that future observations
of nickel, iron and cobalt lines several days after the burst are an 
important test of 
this model.  Since the decay of Nickel-56 proceeds by electron capture, 
observation of a reduced rate of nickel decay would be a unique signal
of ionization. 

An interesting consequence of our model is that it would allow us to test
for the presence of a hypernova even if we are observing it off-axis: as
the ejected nickel is faster than the normal ejecta, it is always outside
the general supernova material that explodes more isotropically. This implies
that the X-ray line flux from the volume behind the ionization front will
be scattered more or less isotropically, and be observable from all
directions. 
We can estimate the redshift out to which such a scattered flux would be
visible by noting that the direct flux from the lines is detectable out
to $z\simeq1$. As long as the ionization front has not passed through the
matter, a fraction of order unity of the direct flux will be scattered, and
it will be roughly isotropic after scattering. This means that the flux
towards any off-axis observer will be reduced by a factor 
$4\pi/\Omega_{\rm jet}\sim30$ relative to the line flux for an observer
with a viewing direction within the jet. The distance out to which we can
see the scattered flux is less than the maximum distance for the direct
flux by the square root of that factor, i.e., corresponds to $z\sim0.2$.
For much closer hypernova suspects, out to a few Mpc distance,
even the $\gamma$-ray flux from the decaying nuclei could be observed with
INTEGRAL. In both cases, detection would imply a very unusual event, likely
a hypernova, since normally the iron-group elements produced in a 
supernova remain deeply hidden in the ejecta for a long time before the
overlying layers become optically thin.

If we were to make observations of delayed nickel decay, it would
lend support to the collapsar model of \grbs.  Since
most of the light from the \grbs  comes in the $\gamma$ rays and in the
afterglow, the central engine is difficult to observe.  The type
of recombination lines predicted here, would, if observed, be a unique
window into the center of the \grb.

\acknowledgments

RAMJW is supported in part by NASA (award no. 21098).


\clearpage


\newpage

\newpage

\clearpage

\end{document}